\documentclass[prl,preprint,aps]{revtex4}

\usepackage{graphicx}
\usepackage{dcolumn}
\usepackage{bm}

\begin{document}

\preprint{APS/123-QED}

\title{Differential Dynamic Microscopy: Probing wave vector dependent dynamics
with a microscope}

\author{Roberto Cerbino$^{(1,2)}$}
\email{roberto.cerbino@unimi.it}
\author{Veronique Trappe$^{(1)}$}
\affiliation{$^{(1)}$Department of Physics, University of Fribourg, Chemin du Mus\'ee 3, CH-1700,
Fribourg, Switzerland}
\affiliation{$^{(2)}$Department of Chemistry, Biochemistry and Medical Biotechnologies,
Universit\'a degli Studi di Milano, I-20133 Milano, Italy}

\begin{abstract}
We demonstrate the use of an ordinary white-light microscope for the study of the q-dependent dynamics of colloidal dispersions. Time series of digital video images are acquired in bright field with a fast camera and image differences are Fourier-analyzed as a function of the time delay between them. This allows for the characterization of the particle dynamics independent on whether they can be resolved individually or not. The characteristic times are measured in a wide range of wavevectors and the results are found to be in good agreement with the theoretically expected values for Brownian motion in a viscous medium.
\end{abstract}

\pacs{Valid PACS appear here}
\maketitle

Microscopy and light scattering are widely used in physics, chemistry,
biology and medical laboratories to access information on the structure
and dynamics of mesoscopic systems. While microscopy gives direct
access to real space images, scattering techniques work in reciprocal
space, where information on the structure and dynamics of the system
is obtained respectively from the angular and time dependence of the
scattered light intensity \cite{chu}. These two complementary techniques
have in general very different experimental requirements. White light
sources are usual choices in microscopy, while a certain degree of
coherence of the illuminating beam is required in scattering experiments;
this is usually achieved by using a laser. In the past many attempts
have been made to build a scattering apparatus based on a microscope;
all of these attempts involved the use of a laser as an illumination
source (see for example Refs. \cite{weitz1,weitz2,axelrod,axelrod2,popescu}
and references therein). In some cases special care was taken to ensure
the capability to perform microscopy and scattering experiments simultaneously,
thereby allowing for a powerful combination of the complementary information
obtained by both techniques \cite{weitz1,weitz2}. More recently,
microscope-based laser Dynamic Light Scattering (DLS) experiments
have been developed to study the dynamic properties of samples of
biological interest, such as living macrophage and red blood cells \cite{axelrod2,popescu}. In practice, due to the intrinsic difficulties in building such instruments,
the use of such techniques has been restricted to those laboratories,
where a sufficient expertise in the realization of optical instrumentation
was at hand.

In this letter, we present a conceptual scheme to interpret and analyze
microscopy images that are obtained from samples containing moving
entities. This technique, which we term Differential Dynamic Microscopy (DDM), does not entail any special experimental
requirements, being based on the use of a standard light microscope with a normal illumination
source and a digital video camera. By using the tools of Fourier Optics \cite{goodman}
we provide the means to access information about the sample dynamics
that are equivalent to the one obtained in multi-angle dynamic light
scattering (DLS) experiments \cite{multiangle}. We test DDM by
analyzing time sequences of microscopy images obtained with an aqueous
dispersion of colloidal particles with diameter 73 nm, well below
the resolution limit of the microscope. Additional tests are performed
with larger particles (diameter 420 nm). Our results are in good
agreement with the predictions from the theory of Brownian motion in a viscous medium,
thereby validating the proposed analysis. We believe that DDM is an interesting complement to the well established DLS \cite{bernepecora} and microscope video tracking \cite{tracking}
techniques.

Our experimental setup simply consists in an inverted microscope (LEICA
DM IRB) equipped with a complementary metal-oxide-semiconductor (CMOS)
camera (IDT X-Stream XS-3, $1280\times1024$ pixels, pixel size=$12\mu m$).
No modifications are made to the microscope. The sample is illuminated
by focusing white light on the sample with a condenser lens (numerical
aperture $0.9$); for detection we use a standard objective (magnification
$M=63\times,$ numerical aperture $NA=0.7$). The effective pixel
size, accounting for the magnification $M$, is then given by $d_{pix}=0.19$$\mu m$,
value smaller than the microscope resolution limit imposed by diffraction.
A capillary tube with rectangular section\textbf{ }(Vitrocom, Inc.)
is filled with an aqueous dispersion of polystyrene
spheres of diameter $73$ nm (Duke Scientific, part number 3070A). The concentration is
1\% by weight fraction, sufficiently low to neglect the effect of
interactions between individual spheres. The thickness of the capillary
tube along the microscope optical axis is $100$$\mu m$ and a region
at the center of the capillary tube is imaged onto the camera sensor.

A sequence of $N=1000$ images is acquired with a sampling rate of $100$
samples/s and with an exposure time of $1$ ms. In Fig.1(a) we show a typical
image: the image appears as a bright background with some dark spots.
These dark spots are dust particles located mainly on the protective
glass of the camera sensor. The signal due to the particles is barely
visible as its contribution is small compared to the large background
signal. However, the background is a static contribution, while the
Brownian motion of the particles causes an incessant rearrangement
of their configuration in space. By subtracting two images grabbed
at different times we can eliminate the static contributions in our
image and bring the displaced particles to light. Examples of difference
images obtained by this subtraction procedure are shown in Fig.1(b)
and 1(c); (b) and (c) are respectively the difference images obtained
by subtracting two images separated in time by $\triangle t=10$ms
and $\triangle t=200$ms. Clearly, the grainy appearance of the difference
images increases with increasing time delay $\triangle t$, revealing
the increasing displacement of the particles.

\begin{figure}
\includegraphics[width=15cm]{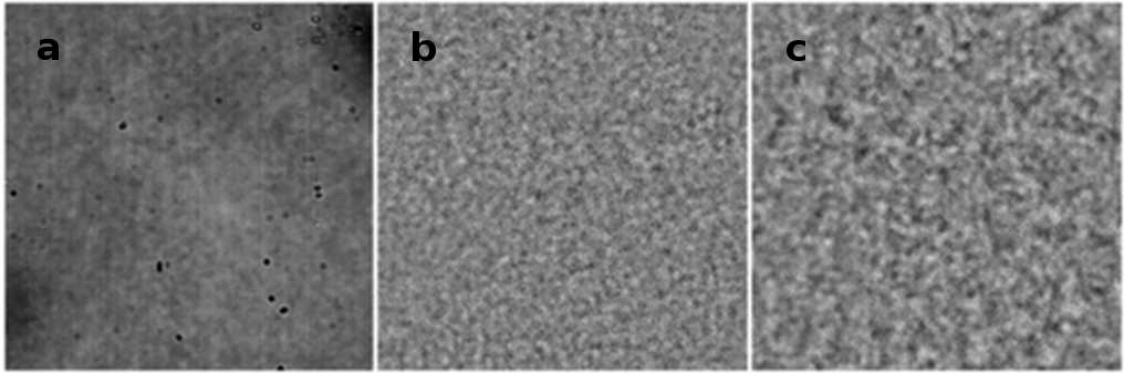}

\caption{(a) Microscope image of an aqueous dispersion of 73
nm particles at a weight fraction of 1\%. The particles are below the resolution of the
microscope and the signal generated by them is very small. (b) Result
of the subtraction of two images taken 10 ms apart in time. The signal
due to the particles is now visible. (c) Result of the subtraction
of two images 200 ms apart in time. The signal has increased. The size
of each panel corresponds to 47 $\mu m$ in the sample.}

\end{figure}

We can quantify this increase by calculating the spatial variance
\begin{equation}
\sigma^{2}(\triangle t)=\int\left|D(x,y;\triangle t)\right|^{2}dxdy,\label{eq:varreal}\end{equation}
of the difference signal $D(x,y;\triangle t)=I(x,y;\triangle t)-I(x,y;0)$
as a function of $\triangle t$. Here, $I(x,y;t)$ is the intensity
obtained in the sensor plane $(x,y)$ at the time $t$, where we define
the acquisition time of the first image to be $t=0$ s. The actual
choice of the reference image is not crucial in the case of Brownian
motion, since the statistical properties of the investigated dynamics
do not change in time. This can be appreciated in the movie clips
shown in \cite{movies}, where the time delay between two subtracted
images is kept constant at $\triangle t=200$ ms. Consequently, for
every $\triangle t$ we calculate the variance $\sigma^{2}(\triangle t)$
by averaging over many statistical realizations of the signal, typically
100, to improve the statistical accuracy. As shown in Fig.2, the
variance $\sigma^{2}(\triangle t)$, increases very rapidly with $\triangle t$
and attains a steady state plateau at long times, when all the particles
have moved far away from their initial position, such that the two
images used to obtain the difference image are statistically uncorrelated.
This dependence of the variance on $\triangle t$ clearly indicates
that difference images can be used to monitor the particle dynamics.
The question then arises whether it is possible to extract quantitative
information on the dynamics of the system from the difference images.
The answer is yes, but its justification requires the use of two steps.

\begin{figure}
\includegraphics[width=15cm]{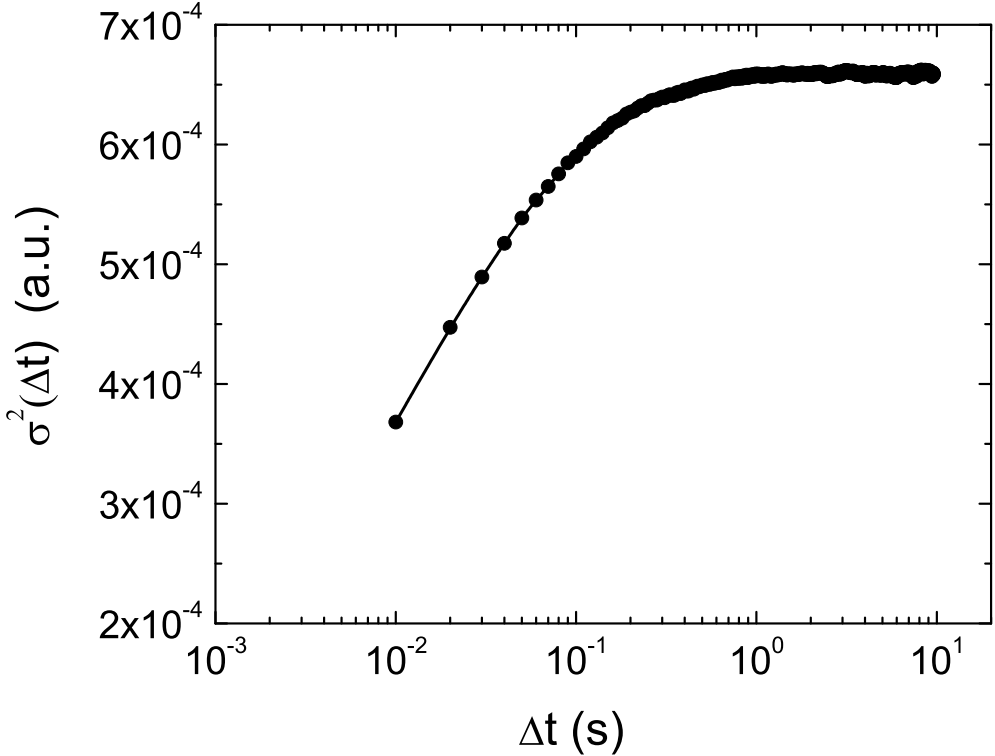}

\caption{The particles rearrangement due to their Brownian motion causes an
increase of the variance $\sigma^{2}(\triangle t)$ of the difference
signal $D(x,y;\triangle t)$. With increasing time delay the two subtracted
images become progressively uncorrelated, such that $\sigma^{2}(\triangle t)$ saturates at large $\triangle t$.}

\end{figure}

The first one is to consider that the variance, defined in Eq.\ref{eq:varreal}
can be calculated in the Fourier space as well. If we define the 2D
Fourier transform of $D(x,y;\triangle t)$ as $F_{D}(u_{x},u_{y};\triangle t)=\int D(x,y;\triangle t)exp\left[-i2\pi\left(u_{x}x+u_{y}y\right)\right]dxdy$,
the use of the Parseval theorem \cite{goodman} guarantees that\begin{equation}
\sigma^{2}(\triangle t)=\int\left|F_{D}(u_{x},u_{y};\triangle t)\right|^{2}du_{x}du_{y}.\label{eq:varfour}\end{equation}
The equivalence of Eq.\ref{eq:varreal} and \ref{eq:varfour}
shows that the total energy content of $D$ in real space is the same
as the one of $F_{D}$ in Fourier space. However, in Fourier space
it is possible to isolate every Fourier component and study the growth
of its amplitude as a function of $\triangle t$.

To clarify whether this mode decomposition provides insight about
the dynamics of the system we need to go through the second step.
This step consists in realizing that a Fourier component of the microscope
image is associated in a simple way to a scattering angle and in turn
to a scattering wavevector. According to Abbe's theory of microscope
image formation \cite{goodman,abbe}, the object to be imaged on the
camera sensor can be thought as the superposition of different Fourier
components characterized by a spatial frequency $u_{obj}$. Each frequency
component of the object acts as a grating and diffracts light at an
angle $\vartheta=sin^{-1}(\lambda u_{obj})$ with respect to the microscope
optical axis, where $\lambda$ is the wavelength of light. On the
camera sensor, the diffracted plane wave will produce a sinusoidal
fringe pattern, characterized by a spatial frequency $u_{det}=sin(\vartheta)/\lambda=u_{obj}$,
independent of the wavelength $\lambda$. This guarantees that we
can associate in a unique way, a spatial frequency $u_{det}$ on the
detector to a scattering angle $\vartheta$.

\begin{figure}
\includegraphics[width=15cm]{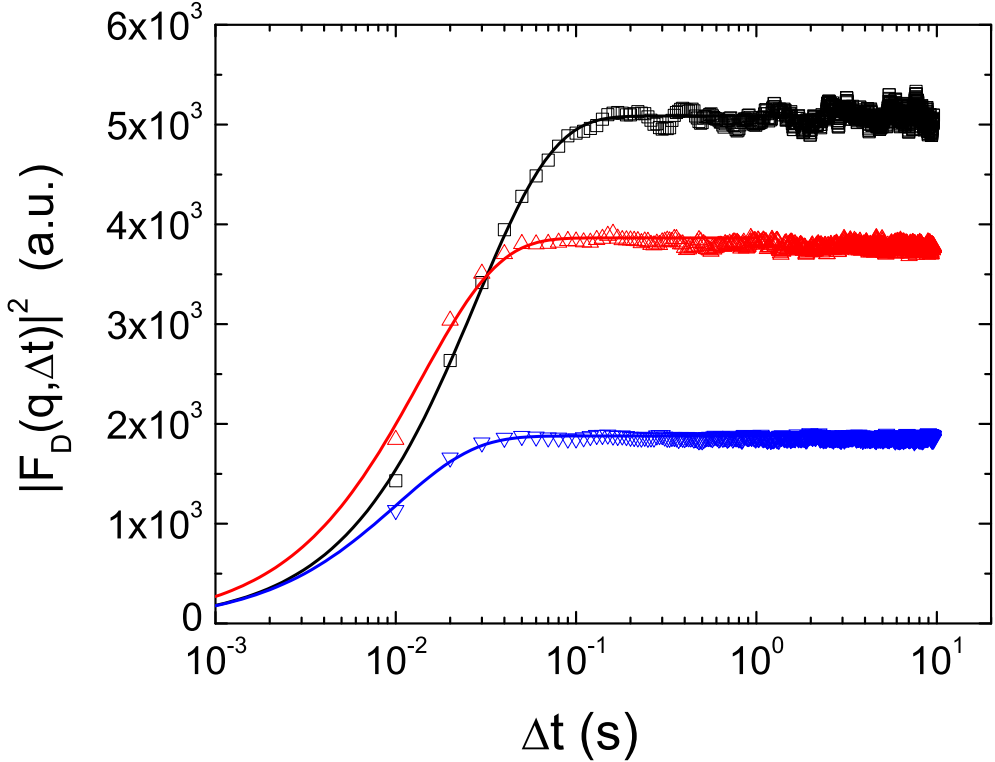}

\caption{Growth of $\left|F_{D}(q;\triangle t)\right|^{2}$
with $\triangle t$ for three values of q: $q=2.6$$\mu m^{-1}$(black squares) , $q=3.9$$\mu m^{-1}$(red
up-triangles) and $q=4.5$$\mu m^{-1}$(blue down-triangles). The continuous lines are fits of the data to Eq.\ref{eq:structure}.}

\end{figure}

To investigate the angular dependent dynamics of our system we thus
use the Fourier power spectrum $\left|F_{D}(u_{x},u_{y};\triangle t)\right|^{2}$,
which we simply calculate by applying a Fast Fourier Transform \cite{bracewell}
to the difference images. For isotropic samples, like the one investigated,
the expectation value of $\left|F_{D}(u_{x},u_{y};\triangle t)\right|^{2}$
is rotationally invariant in the $(u_{x},u_{y})$ plane. This allows
us to perform azimuthal averages and to treat the one-dimensional
power spectrum $\left|F_{D}(u;\triangle t)\right|^{2}$, where $u=\sqrt{u_{x}^{2}+u_{y}^{2}}$.
For easy comparison with scattering experiments we use wavevectors
$q=2\pi u$ instead of spatial frequencies $u$. Processing the data
at several $\triangle t$ enables us to report the one-dimensional
power spectrum $\left|F_{D}(q;\triangle t)\right|^{2}$ as a function
of the time delay, as shown for three values of $q$ ($q=2.6$$\mu m^{-1}$,
$q=3.9$$\mu m^{-1}$ and $q=4.5$$\mu m^{-1}$) in Fig.3. As for
the variance $\sigma^{2}(\triangle t)$ we find that $\left|F_{D}(q;\triangle t)\right|^{2}$
increases rapidly with $\triangle t$ to then saturate at longer
time delays. The cross-over to the plateau value systematically shifts
to smaller $\triangle t$ as the wavevector $q$ increases, denoting
a $q$-dependence in the characteristic time of the system.

For a quantitative description of this behavior we consider that fluctuations in the intensity of the original images are due to concentration fluctuations in the sample. For Brownian diffusion it is well known that every Fourier concentration mode decays exponentially in time, $exp(-\triangle t/\tau(q))$, with a characteristic time $\tau(q)=1/D_{m}q^{2}$, where $D_{m}$ is the mass diffusion coefficient of the particles \cite{bernepecora}. For the analysis of the difference images it is then easy to show that \begin{equation}
\left|F_{D}(q;\triangle t)\right|^{2}=A(q)[1-exp(-\triangle t/\tau(q))]+B(q).\label{eq:structure}\end{equation}
This equation contains two main contributions: the term $B(q)$ is due to the power spectrum of the camera noise and is present even in absence of the particles. The term $A(q)[1-exp(-\triangle t/\tau(q))]$ describes the contribution associated to the particles. A quantitative description of the term $A(q)$ requires the knowledge of the exact relationship between intensity and concentration fluctuations, which is beyond the scope of this paper. However once $q$ is fixed, we can treat $A(q)$ and $B(q)$ as simple adjusting parameters and extract the characteristic time $\tau(q)$ by simply considering the $\triangle t$-dependence of $|F_{D}(q;\triangle t)|^{2}$. This is very similar to what has been done in Ref. \cite{fab1} where an expression similar to Eq.\ref{eq:structure} has been proposed to analyze sequences of shadowgraph images for the characterization of nonequilibrium concentration fluctuations in a binary mixture.

\begin{figure}
\includegraphics[width=15cm]{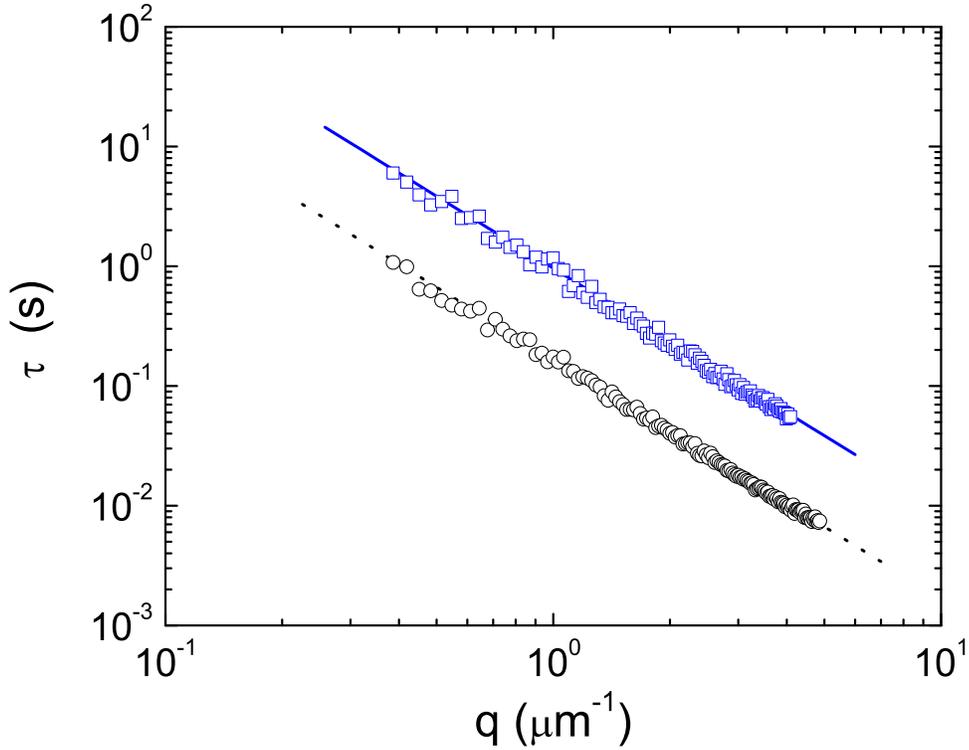}

\caption{Characteristic decay time $\tau$ versus the wavevector $q$ for the
73 nm (black circles) and the 420 nm (blue squares) particles. Fitting of the data gives respectively the values
$D_{m}=6.2\pm0.3$ $\mu m^{2}/s$ and $D_{m}=1.1\pm0.2$ $\mu m^{2}/s$. The
lines are theoretical predictions calculated by using the Stokes-Einstein
relation with no adjustable parameters.}

\end{figure}

Thus, we fit our data to Eq.\ref{eq:structure} to obtain $\tau(q)$, which we report in Fig.4 for $q$ ranging from $0.4$ $\mu m^{-1}$ to $5$ $\mu m^{-1}$ (black open circles). On the same graph the dotted black line denotes the expected behavior calculated by using the formula $\tau(q)=1/D_{m}q^{2}$, where $D_{m}$ is estimated using the Stokes-Einstein
relation $D_{m}=k_{B}T/(3\pi\eta d)=5.98$ $\mu m^{2}/s$; $k_{B}$
is the Boltzmann constant, $T$ the absolute temperature, $\eta$
the solvent viscosity and $d$ the diameter of the particles. The
experimental results are in good agreement with the theoretically
expected values over more than one decade in $q$, showing that DDM is a reliable means to determine the dynamical properties
of the sample. To our knowledge this result provides the first experimental
evidence that white light, bright-field microscopy can be used to
monitor the $q$-dependent dynamics of particles.

\begin{figure}
\includegraphics[width=15cm]{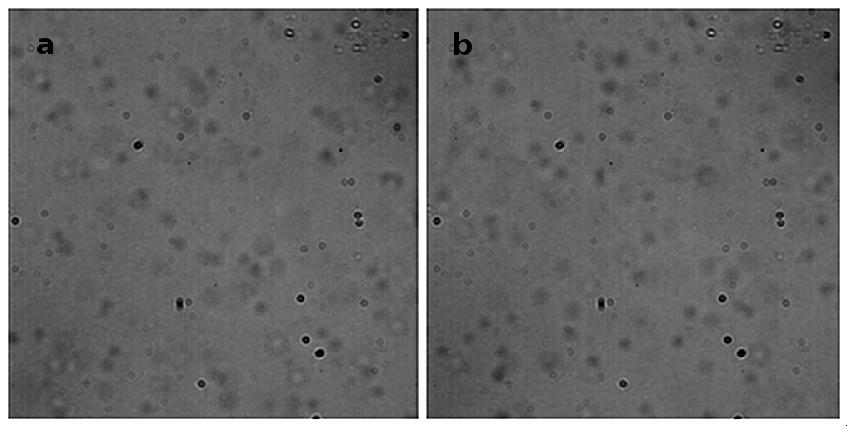}

\caption{Microscope images taken 10 s apart. The side of each panel corresponds
to 47$\mu m$ in the sample. The sample is a colloidal dispersion
of 420 nm particles. The more visible objects are dust particles on
the CMOS sensor which are also present in Fig.1(a), grabbed with
another sample. The particles are the less visible objects that change
position between image (a) and (b). The smearing effect due to the
finite point spread function of the microscope is visible.}

\end{figure}

We also perform experiments by using particles with a diameter of $420$ nm, value comparable to the point spread function of the microscope. In Fig.5 we present two images of this sample separated in time by
10 s. In contrast to the previous case the particles are fairly visible,
though the static signal due to dust along the optical path still
represents a significant contribution. However, the subtraction procedure
described above can still be applied to visualize the particles motion and to eliminate the static noise, as shown in the movie clip presented in \cite{movies}. The image sequence is analyzed as described
above and the results are presented in Fig.4 as blue open squares. A good agreement with the theoretically expected behavior ($D_{m}=1.04\mu m^{2}/s$,
continuous blue line) is also found in this case, demonstrating that DDM is capable of monitoring the dynamics of particles that
can be either resolved individually or not. This suggests that DDM could be a valid complement to tracking techniques \cite{tracking} for relatively big particles, with the advantage of being also applicable to concentrated systems, where tracking becomes difficult, and to fluctuating systems which are not composed by particles. Typical examples are membranes, interfaces or liquid crystals. For small particles DDM can be subsidiary to DLS experiments \cite{bernepecora}, because it can operate at small wavevectors where stray light makes the use of DLS notoriously difficult. The range of accessible time-scales is limited by the inverse of the acquisition frame rate on the lower side and by the overall duration of the experiment on the higher end.

A possible extension of this work could be the investigation of dynamically heterogeneous systems, where a space
resolved investigation of the dynamics is of fundamental importance \cite{cipram}.
Our technique could also be applied to absorbing samples and to fluorescence
microscopy, which is widely used for the investigation of biological
samples \cite{fluormicr}. Finally, we believe that DDM
can be profitably used to perform microrheology experiments as pioneered
in Ref. \cite{mason}.

\begin{acknowledgments}
The authors thank M. Carpineti, F. Giavazzi and A. Vailati for stimulating
discussions and for critical reading of the manuscript. T. Bellini,
P. Cicuta, F. Croccolo, M. Giglio, R. Piazza, C. Takacs and D. Weitz are thanked
for fruitful discussions. R.C. acknowledges financial support from
the European Union (Marie Curie Intra-European Fellowship, contract
EIF - 038772) and V.T. financial support from the Swiss National Science
Foundation.
\end{acknowledgments}

\end{document}